\documentclass[final,times,twocolumn]{elsarticle}

\usepackage[hidelinks]{hyperref}
\graphicspath{{figs/}{.}}
\usepackage{textcomp}
\usepackage{siunitx}
\newcommand{\electron}{\rm e^{-}}
\usepackage{textgreek}
\usepackage{paralist}
\usepackage{amssymb}
\usepackage{paralist, sectsty, textcase}

\journal{Nuclear Instrumentation and Methods A}

\begin{document}

\begin{frontmatter}



\title{Megapixels @ Megahertz – 
The AGIPD High-Speed Cameras for the European XFEL}


\author[1]{Aschkan Allahgholi}		
\author[1]{Julian Becker}		 	
\author[1]{Annette Delfs}			
\author[2]{Roberto Dinapoli}		
\author[1]{Peter Göttlicher}		
\author[1,5]{Heinz Graafsma}		
\author[2]{Dominic Greiffenberg}	
\author[1]{Helmut Hirsemann}		
\author[1]{Stefanie Jack}			
\author[1]{Alexander Klyuev}		
\author[4]{Hans Krüger}				
\author[1]{Manuela Kuhn}			
\author[1]{Torsten Laurus}			
\author[1]{Alessandro Marras}		
\author[2]{Davide Mezza}			
\author[2]{Aldo Mozzanica}			
\author[1]{Jennifer Poehlsen}		
\author[1]{Ofir Shefer Shalev}		
\author[1]{Igor Sheviakov}			
\author[2]{Bernd Schmitt}			
\author[3]{Jörn Schwandt}			
\author[2]{Xintian Shi}				
\author[1]{Sergej Smoljanin}		
\author[1]{Ulrich Trunk\corref{cor1}}\ead{ulrich.trunk@desy.de}
\author[2]{Jiaguo Zhang}			
\author[1]{Manfred Zimmer}			

\cortext[cor1]{Corresponding author}

\address[1]{Deutsches Elektronensynchrotron - DESY, Hamburg, Germany}
\address[2]{Paul Scherrer Institut - PSI, Villigen, Switzerland}
\address[3]{Universität Hamburg, Hamburg, Germany}
\address[4]{Universität Bonn, Bonn, Germany}
\address[5]{Mid-Sweden University, Sundsvall, Sweden}

\begin{abstract}
The European XFEL is an extremely brilliant Free Electron Laser Source with a very demanding pulse structure: trains of 2700 X-Ray pulses are repeated at \SI{10}{\hertz}. The pulses inside the train are spaced by \SI{220}{\nano\second} and each one contains up to \SI{e12}{photons} of \SI{12.4}{\kilo\electronvolt}, while being $\rm \leq \SI{100}{\femto\second}$ in length.
AGIPD, the Adaptive Gain Integrating Pixel Detector, is a hybrid pixel detector developed by DESY, PSI, and the Universities of Bonn and Hamburg to cope with these properties. 

It is a fast, low noise integrating detector, with single photon sensitivity (for $E_{\gamma} \gtrapprox \SI{6}{\kilo\electronvolt}$) and a large dynamic range, up to \num{e4} photons at \SI{12.4}{\kilo\electronvolt}. This is achieved with a charge sensitive amplifier with 3 adaptively selected gains per pixel. 352 images can be recorded at up to \SI{6.5}{\mega\hertz} and stored in the in-pixel analogue memory and read out between pulse trains. The core component of this detector is the AGIPD ASIC, which consists of $64 \times 64$ pixels of $\SI{200}{\micro\meter} \times \SI{200}{\micro\meter}$. Control of the ASIC's image acquisition and analogue readout is via a command based interface. FPGA based electronic boards, controlling ASIC operation, image digitisation and \SI{10}{GE} data transmission interface AGIPD detectors to DAQ and control systems.

An AGIPD \SI{1}{\mega pixel} detector has been installed at the SPB\footnotemark
experimental station in August 2017, while a second one is currently commissioned for the MID\footnotemark
 endstation. A larger (\SI{4}{\mega pixel}) AGIPD detector and one to employ Hi-Z sensor material to efficiently register photons up to $E_{\gamma} \approx \SI{25}{\kilo\electronvolt}$ are currently under construction.

\end{abstract}


\end{frontmatter}
\setcounter{footnote}{0}
\stepcounter{footnote}
\footnotetext{Single particles, Clusters and Biomolecules}
\stepcounter{footnote}
\footnotetext{Materials Imaging and Dynamics}

\section{The European XFEL}
\label{Sect:EuXFEL}

\begin{figure}[!tb]
\includegraphics[width=\columnwidth]{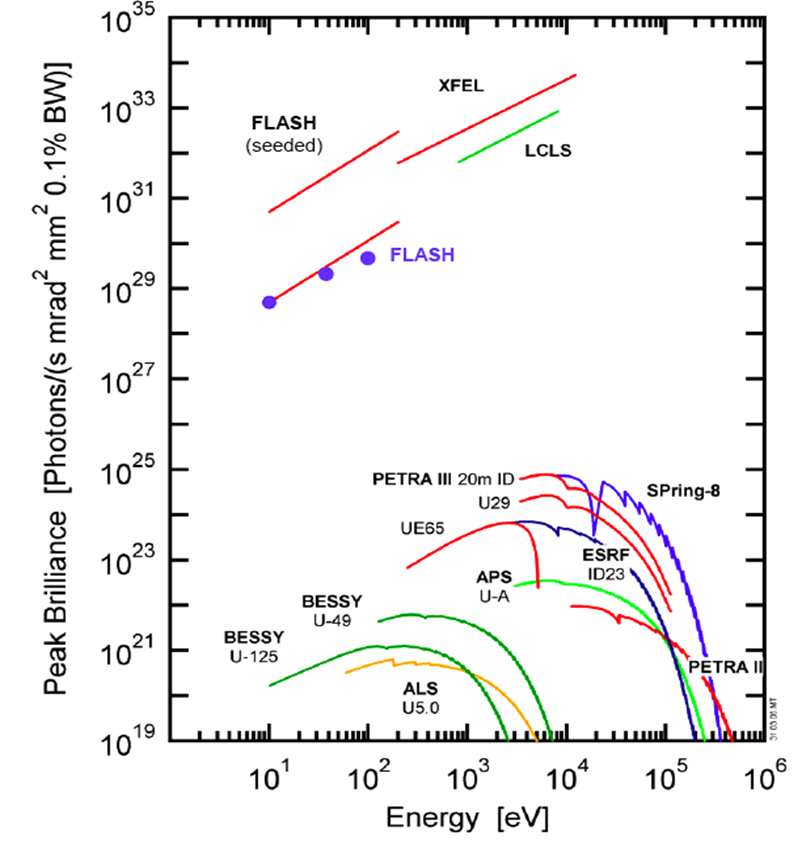}
\caption{Brilliance of FELs and Synchrotron sources.}
\label{Brilliance}
\end{figure}

\begin{figure}[!tb]
\includegraphics[width=\columnwidth]{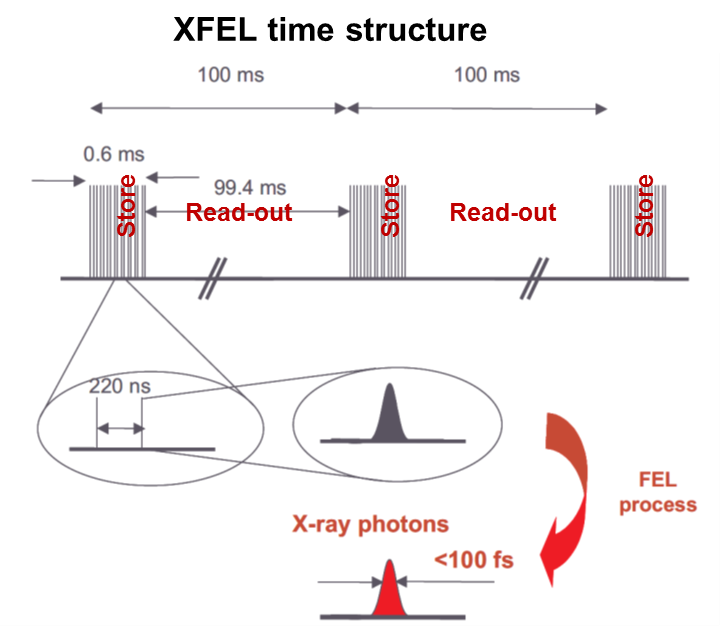}
\caption{Time structure of the European XFEL source.}
\label{XFELTiming}
\end{figure}

The European X-Ray Free Electron Laser (XFEL) \citep{1, EXFEL} in Hamburg is currently the most brilliant X-Ray source (fig.~\ref{Brilliance}) in the world. It provides extremely focused, fully coherent X-Ray pulses. Trains of 2700 of these pulses are repeated at \SI{10}{Hz}. The pulses inside the train are spaced by \SI{220}{\nano\second} and contain up to $\num{e12} \times \SI{12.4}{\kilo\electronvolt}$ \footnote{\SI{12.4}{\kilo\electronvolt} is the maximum fundamental photon energy of the SASE1 and SASE2 undulators at the European XFEL, providing beam for the AGIPD-equiped instruments.} photons each, while being $\le \SI{100}{\femto\second}$ in length (fig.~\ref{XFELTiming}). The high intensity per pulse will allow recording diffraction patterns of single molecules or small crystals in a single shot. As a consequence 2D detectors have to cope with a large dynamic range, requiring single photon sensitivity e.g. for single molecule imaging, and registering more than \num{e4}\si{photons\per pixel} in the same image for the case of liquid scattering or intense Bragg Spots\footnote{Droplets and ice crystals are frequently encountered in gas and liquid jet sample delivery setups}. 
AGIPD is one of three detectors developed to cope with the timing requirements of the European XFEL\citep{2}. The other two are the Large Pixel Detector (LPD) \cite{LPD} with $\SI{500}{\micro\meter} \times \SI{500}{\micro\meter}$ pixel size and \SI{1}{\mega pixel} installed,  and DSSC \cite{DSSC}, a \SI{1}{\mega pixel} detector with $\SI{200}{\micro\meter} \times \SI{230}{\micro\meter}$ hexagonal pixels, based on silicon drift diodes or Depfet sensors with Signal Compression (hence the acronym), currently under development. While LPD is targeted at the same energy range as AGIPD, it differs from it by a lower spatial and energy resolution, but covers a much bigger area and operates in ambient. DSSC like AGIPD operates in vacuum, but can detect lower photon energies - on the expense of a nonlinear response and a lower frame rate at low energies. 

Future upgrades of the European XFEL planned for the $\rm 2^{nd}$ half of the 2020ies include continuous wave (CW) operation at a pulse rate of $\approx \SI{100}{\kilo\hertz}$ and a so-called \textsl{long pulse} mode at $\lessapprox \SI{200}{\kilo\hertz}$ with \SI{500}{\milli\second} bursts repeating at \SI{1}{\hertz}\cite{CW_LP}. At these rates moving or exchanging solid samples during a burst becomes feasible, which would reduce the radiation damage of samples considerably and thus enable european X FEL for new classes of experiments and materials. To exploit these modes new detectors with a different readout architecture are needed, since AGIPD hardware can\footnote{The firmware for the readout system would have to send commands to the ASIC in a different order} only operate up to $\approx \SI{16}{\kilo\hertz}$ 
CW frame rate.

\section{The AGIPD Detector}
\label{Sect:AGIPDdet}
AGIPD \citep{AGIPD,3,4,AA_IWORID15} is a hybrid pixel detector, developed by DESY, the Paul Scherrer Institute and the Universities of Bonn and Hamburg to meet the requirements for the use at the European XFEL. 
It consists of \SI{500}{\micro\meter} thick Silicon sensors, manufactured by SINTEF, Hamamatsu and ADCVACAM, to provide a high efficiency ($\rm QE \ge 98 \%$) up to \SI{12.4}{\kilo\electronvolt} photon energy. While the sensors efficiently sheild the ASICs underneath from radiation, the sensors themselves will collect a substantial amount of radiation dose during their expected lifetime. Special design measures have been taken to mitigate radiation damage and its effects \cite{Sensor}. Each sensor features $512 \times 128$ pixels, which are $\SI{200}{\micro\meter} \times \SI{200}{\micro\meter}$ in size and pitch as a good compromise. $8 \times 2$ AGIPD ASICs are bump-bonded to each sensor for readout. Larger detector systems of 1- or 4-megapixels are composed by tiling these sensors\footnote{The minimum distance of the individual sensor tiles are \SI{0.4}{\milli\meter} in horizontal and \SI{3.4}{\milli\meter} in vertical direction respectively.}.  The high photon flux together with single photon sensitivity require the detector to operate in vacuum, in order to prevent the intense beam from interacting with ambient air or exit windows, which would cause a huge background. In turn a vacuum vessel is an integral part of the detectors. The \SI{1}{\mega pixel} detectors at the SPB and MID experimental stations have the $\rm 4 \times 4$ sensor modules arranged on four movable quadrants\footnote{The AGIPD \SI{1}{\mega pixel} detector for the HIBEF endstation features a different layout}. These can be arranged to form a hole for the direct beam to prevent it from hitting detector components and inflicting damage to the system. 
The intense beam and the experiments envisioned also require a huge dynamic range, which can reach \SI{e4}{photons\per pixel\per image} at \SI{12.4}{\kilo\electronvolt}. To cope with this, AGIPD adaptively lowers the sensitivity of the preamplifiers, independently for each pixel in two steps.
Furthermore the system has to comply with the rather inconvenient time structure of the European XFEL. Since it is not possible to read out an image within \SI{220}{\nano\second}, the detector has to record as many images as possible during a pulse train and read these out during the 99.4\,ms gap in-between the trains. 

\section{The AGIPD readout ASIC}
\label{Sect:AGIPDasic}	
\begin{figure*}[!tb]
\includegraphics[width=\textwidth]{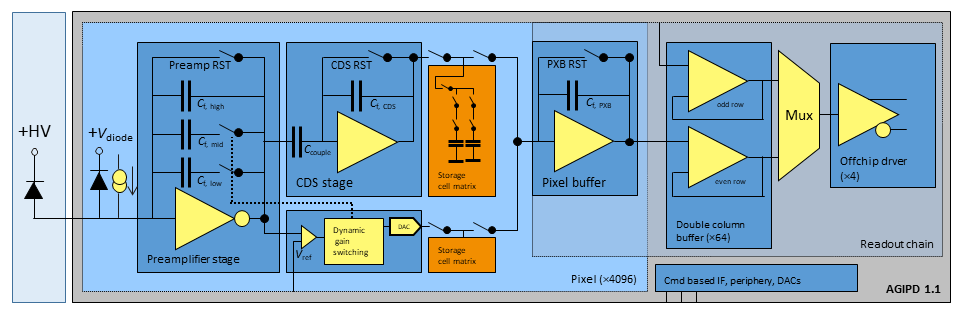}
\caption{Circuit schematic of the AGIPD 1.1 ASIC.}
\label{AGIPD_schema}
\end{figure*}

The core functionality of the AGIPD detector is implemented in the readout ASIC \citep{5, AGIPD_Shi, DM_NIM16}. It is manufactured in IBM/Global Foundries cmrf8sf (\SI{130}{\nano\meter}) technology and contains $\rm 64 \times 64$ pixels of $\SI{200}{\micro\meter} \times\SI{200}{\micro\meter}$.  The circuit in each pixel (fig.~\ref{AGIPD_schema}) contains a charge sensitive preamplifier based on an inverter core  with threefold switchable gain. This gain switching is implemented by adding capacitors of \SI{3}{\pico\farad} and \SI{10}{\pico\farad} to the initial preamplifier feedback of \SI{60}{\femto\farad}. A discriminator is connected to the preamplifier output and triggers an adaptive gain selection, whenever this output exceeds a selected threshold. A correlated double sampling (CDS) stage removes reset and attenuates low-frequency noise components from the preamplifier output \citep{Buttler}, such that the detection of single photons\footnote{down to $\rm \approx \SI{6}{\kilo\electronvolt}$ with SNR $\rm \ge \SI{5}{\sigma}$} is feasible. 
The output of the CDS, as well as the selected gain is sampled in a capacitor based analogue memory for 352 images, which occupies about \SI{80}{\percent} of a pixel's area. It is based on n-FET in n-well capacitors and dual PMOS switches for radiation tolerance and low leakage, since the analogue signal must not deteriorate during several \SI{10}{\milli\second} of storage time. For readout each pixel features a $\rm 2^{nd}$ charge sensitive buffer. Last but not least each pixel features two sources of electrical stimuli: A constant curren source and a pulsed capacitor, which can be connected to the preamplifier input. A command based interface and control circuit provides random access to the memory and controls the row-wise readout of the data via multiplexers to four differential analogue ports. The random access scheme allows overwriting image data within a bunch train, which is compliant with the European XFEL's vetoing schema to maximise efficiency. The data of individual 'amplitude' and 'gain' frames will be combined to single images by the readout electronics\footnote{The current firmware of the AGIPD \SI{1}{\mega pixel} detectors at SPB and MID reads 300 separate 'amplitude' and 'gain' frames, which are combined offline.}. The power consumption of an ASIC is typically around \SI{1.0}{\ampere} at \SI{1.5}{\volt}. To ensure sufficient radiation tolerance of the ASIC, individual components and building blocks have been tested for doses up to several \SI{10}{\kilo\gray}, e.g. in \cite{UT_IEEE12}.

\subsection{ASIC Performance}
\label{Sect:AGIPDperf}
 
\begin{figure}[!hb]
\centering
\includegraphics[width=\columnwidth]{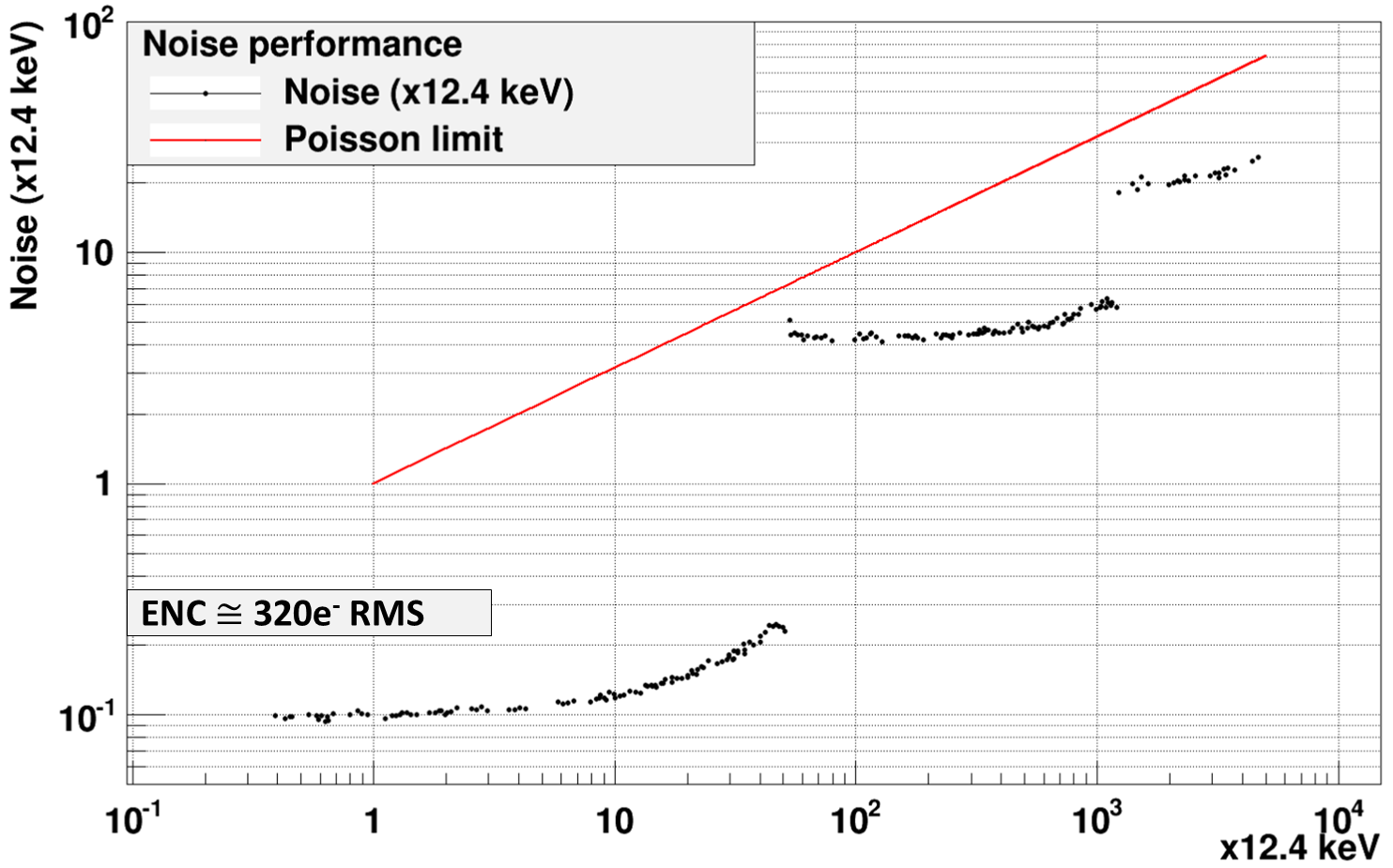}
\caption{Noise performance of AGIPD \citep{AA_IWORID15}. An IR laser was used to access the different gains. Fluctuations of this stimulus add an amplitude-dependent contribution to the measured ENC.}
\label{AGIPD_noise}
\end{figure}

\begin{figure}[!hb]
\centering
\includegraphics[width=\columnwidth]{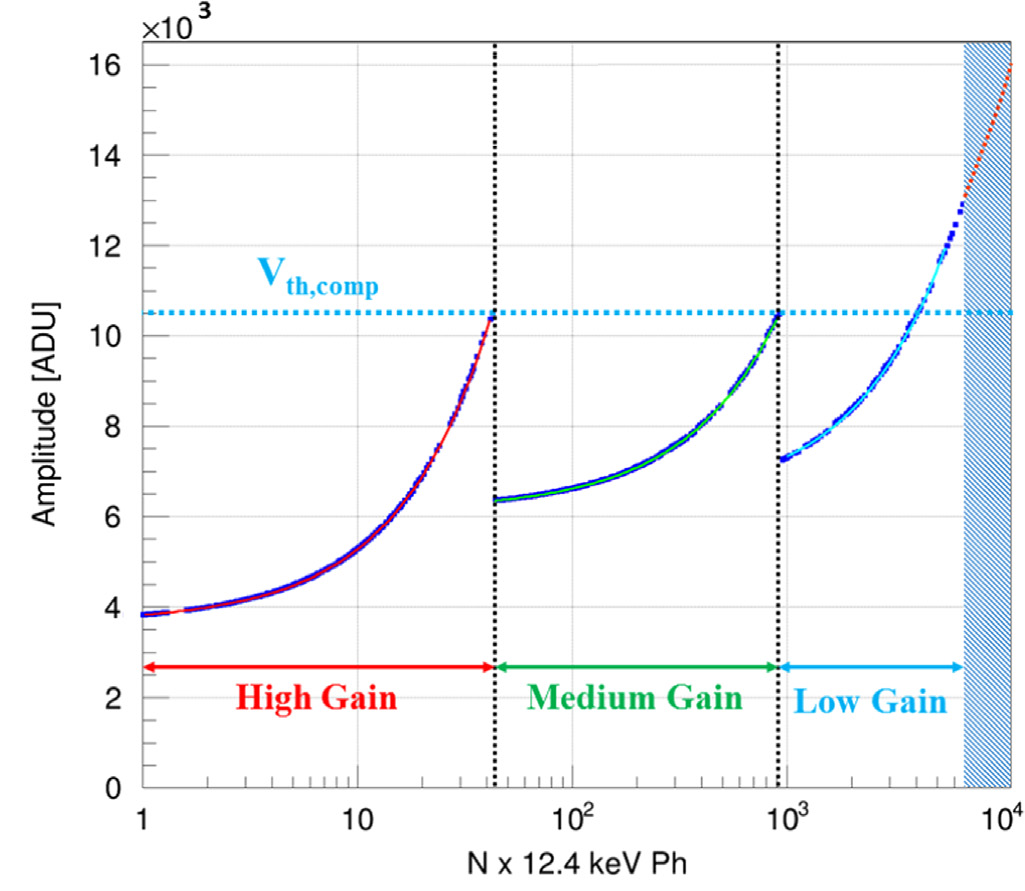}
\caption{Transfer characteristics of AGIPD, including gain switching \citep{DM_NIM16}. The solid lines are linear fits to the corresponding gain setting.}
\label{AGIPD_dr}
\end{figure}

\begin{figure}[!tb]
\centering
\includegraphics[width=\columnwidth]{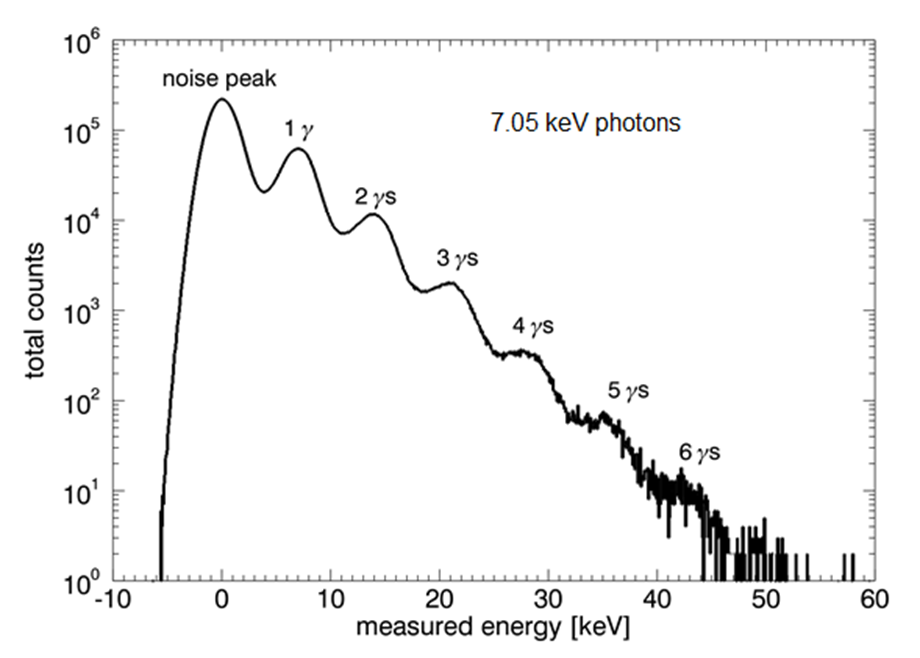} 
\caption{Spectrum of 7keV photons measured by AGIPD (data from Petra III).}
\label{AGIPD_fluorescence}
\end{figure}

All AGIPD systems currently in use are based on the AGIPD 1.1 version\footnote{A breif description of earlier versions and prototypes of the ASIC can be found in \cite{UT_IEEE12}.} of the ASIC, which has been thoroughly characterised electrically and with a sensor. Since the medium and low gain settings are not accessible with lab X-Ray sources, an IR laser has been used for characterisation. This way an ENC\footnote{Equivalent Noise Charge} of  $\rm \approx 320 e^{-}$ was measured\footnote{Sacrificing dynamic range by selecting the high gain of the CDS, \mbox{$\rm ENC \approx \SI{240}{e^{-}}$} can be reached.}, as fig.~\ref{AGIPD_noise} shows. This value corresponds to a signal to noise ratio $>\SI{10}{ \sigma}$ for a single \SI{12.4}{\kilo\electronvolt} photon in the high gain \citep{6}. Fig.~\ref{AGIPD_noise} furthermore shows, that the noise in all gain settings is always less than the limit imposed by the Poisson statistics intrinsic to the discrete nature of the impinging photons. \\
The dynamic range was measured up to $\SI{344e6}{e^{-}}$  or \SI{e4}{photons} of \SI{12.4}{\kilo\electronvolt}, as fig.~\ref{AGIPD_dr} shows. The linear fits in fig.~\ref{AGIPD_dr} also confirm a nonlinearity better than \SI{0.44}{\percent} up to \SI{5e3}{photons} of \SI{12.4}{\kilo\electronvolt}. The sensitivity to single photons is confirmed by Fig. \ref{AGIPD_fluorescence}.

\section{AGIPD Calibration Procedure}
\label{Sect:Calibration}
Calibrating a detector system with multiple gains is a non-trivial task, since the medium and low gain settings of the preamplifier are not easily accessible with radiation sources. While the intensity of the direct beams at FELs and synchrotrons might be sufficient to reach these, the statistics required for the necessary precision and the associated time and radiation load, which is expected to vastly exceed that of experiments, render this approach impractical. Instead a hybrid approach based on  fluorescence photons and electrical stimuli provided by the in-pixel constant current source (c.f. sect.\ref{Sect:AGIPDasic}) is used:
\begin{compactitem}
\item Calibration of the preamplifiers high gain with fluorescence photons from Copper or Molybdenum targets.
\item Scan of the preamplifiers integration time (from the nominal \SI{150}{\nano\second} to several \SI{}{\micro\second}) using the in-pixel current source to determine the sensitivity of the medium and low gain settings relative to the high gain.
\end{compactitem}
A big advantage of this approach is the independence from the absolute value of the current, i.e. of variations in the ASIC's manufacturing process or sensor leakage currents, whereas the indirect nature of the procedure and the long \textsl{lever arm} are mitigated by the Poissonian nature of photons, which relaxes requirements for higher intensities.

Using the DAQ system at the European XFEL, data for a full calibration can be acquired within less than \SI{2}{\hour}, i.e. this can be conveniently performed before and after user experiments. From this data the 3 offsets (\textsl{baseline}) and gains have to be calculated. In addition gain and offset also depend on the size of the memory's capacitors. This leads in total to $\rm (3\, gains + 3\, offsets) \times 352\, storage\; cells \times 1048576\, pixels = 2.219\cdot 10^{9}\, constants$.  Calculating these constants requires $\rm \leq \SI{4}{\hour}$ on the DESY \textsl{Maxwell} computer cluster \cite{DESY_Maxwell}, and no further reduction of constants by disentanglement of the contributions from frontend and memory are performed.
The calculation of the two threshold levels required to re-digitise the the gain information is also included in the calibration process. Remaining uncertainties in determining the gain even after calibration led to the further improved AGIPD 1.2 readout ASIC. 

\section{AGIPD Detector Systems at SPB \& MID}
\label{Sect:AGIPDspb}

The image planes of all AGIPD cameras are composed from \textsl{Frontend Modules} (FEM). These are constructed from Silicon sensor tiles, 
to which $8 \times 2$ AGIPD ASICs are bump-bonded by means of Sn/Ag or Sn/Pb bumps. The sensor tiles are 
p-on-n type with implant sizes and metal overhangs specially tailored to avoid \textsl{dead pockets}, where charges would be trapped and contribute to radiation damage, as well as to facilitate bias voltages up to \SI{900}{\volt} \citep{Sensor}. The latter is required to overcome so called \textsl{plasma effects} \citep{Plasma}, when the impinging radiation e.g. on a Bragg spot creates charge carrier densities able to effectively shield of the drift field. For the first user experiments bias voltages of \SI{300}{\volt} or \SI{500}{\volt} were applied. Further parts of a frontend module are an LTCC\footnote{Low Temperature Co-fired Ceramics} carrier board, to which the sensor assemblies are glued and wire bonded and the copper interposer, to which the LTCCs are bolted. 
For the AGIPD \SI{1}{\mega pixel} detectors for SPB and MID four of these frontend modules are attached to a copper cooling block to form a quadrant. This way the temperature of the modules can be lowered to $\approx \SI{0}{\celsius}$. The quadrants of the \SI{1}{\mega pixel} systems at the SPB and MID experimental stations are attached to a wedge-shaped arrangement of linear stages and mounted inside a vacuum vessel, while the image plane sticks out of it. Connected to the experimental chamber vacuum levels down to \SI{e-7}{} have been reached during user operation\footnote{AGIPD can also operated at ambient pressure, but coarse vacuum levels are -- depending on sensor bias -- forbidden by Paschen's law.}. The movable arrangement of the quadrants permits the formation of an arbitrary hole or slot for the direct beam to pass, while the wedge shaped stages allow the driving motors to be mounted outside the vessel -- on the expense of loosing the orthogonality of the translation axes.

\begin{figure}[!tb]
\centering
\includegraphics[width=\columnwidth]{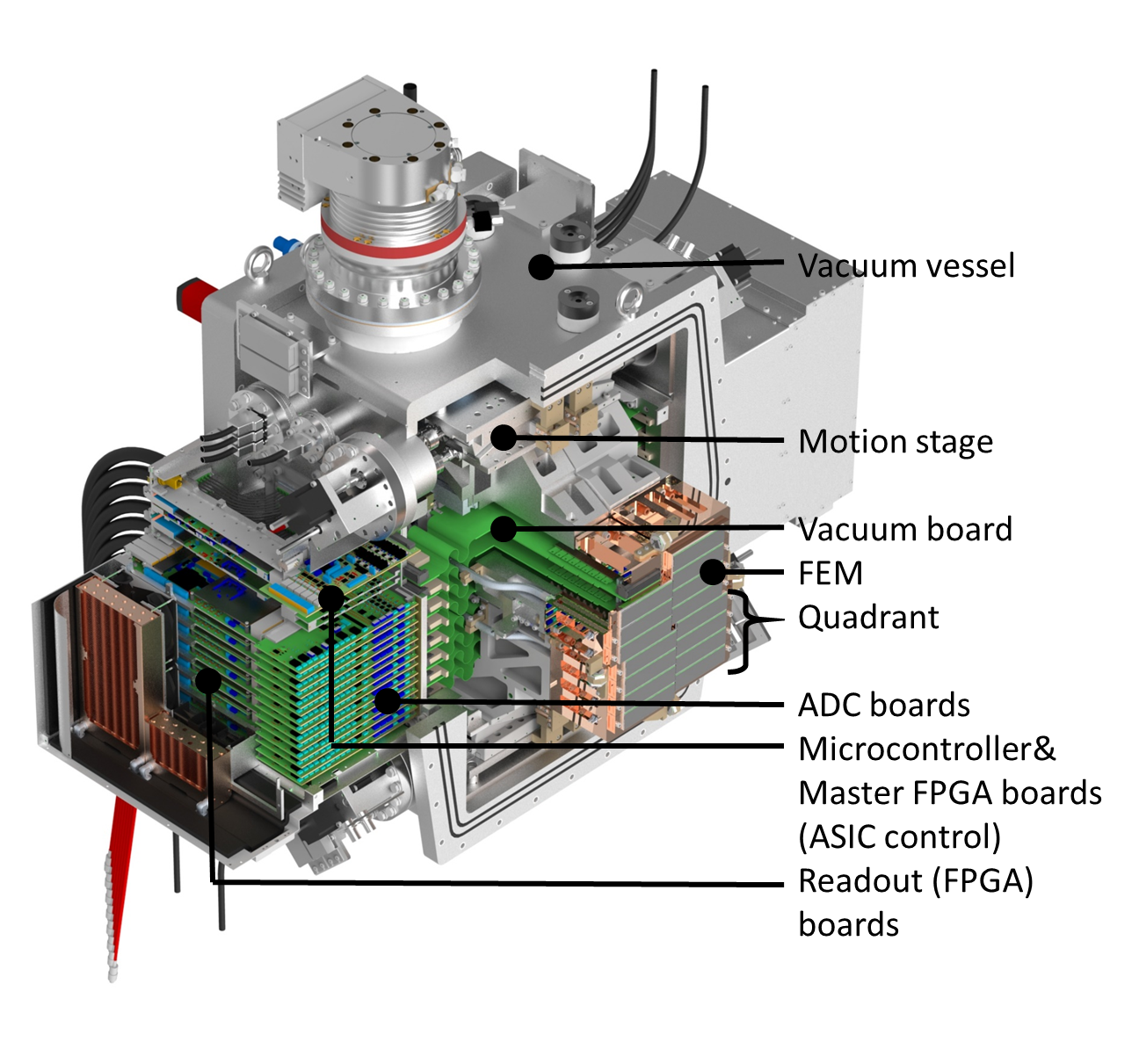} 
\caption{Rendering of the AGIPD \SI{1}{\mega pixel} detectors for the SPB and MID endstations. To show the locations of the components, parts of the vessel and the external housing are cut out.}
\label{AGIPD_rendering}
\end{figure}

The 64 analogue signals of one frontend module are brought outside the detector's vacuum vessel via the \textsl{vacuum board}, a PCB\footnote{Printed Circuit Board} with a flexible section to compensate for the motion of the quadrants, and a vacuum flange formed by another PCB serving as the vacuum barrier. Outside the vacuum vessel PCBs with receiver amplifiers and ADCs provide digitisation of the data with 14\,bit quantisation. This data is then collected by an FPGA\footnote{Field-Programmable Gate Array} and sent on via an 10\,GB optical ethernet link per FEM to the data acquisition system of the European XFEL, which combines the data of individual modules to full frames and passes them on to storage.
A rendering of the \SI{1}{\mega pixel} detectors for the SPB and MID endstations is shown in fig. \ref{AGIPD_rendering}. The power consuption of such a system is about \SI{1.2}{\kilo\watt} excluding cooling.

\begin{figure}[!tb]
\centering
\includegraphics[width=\columnwidth]{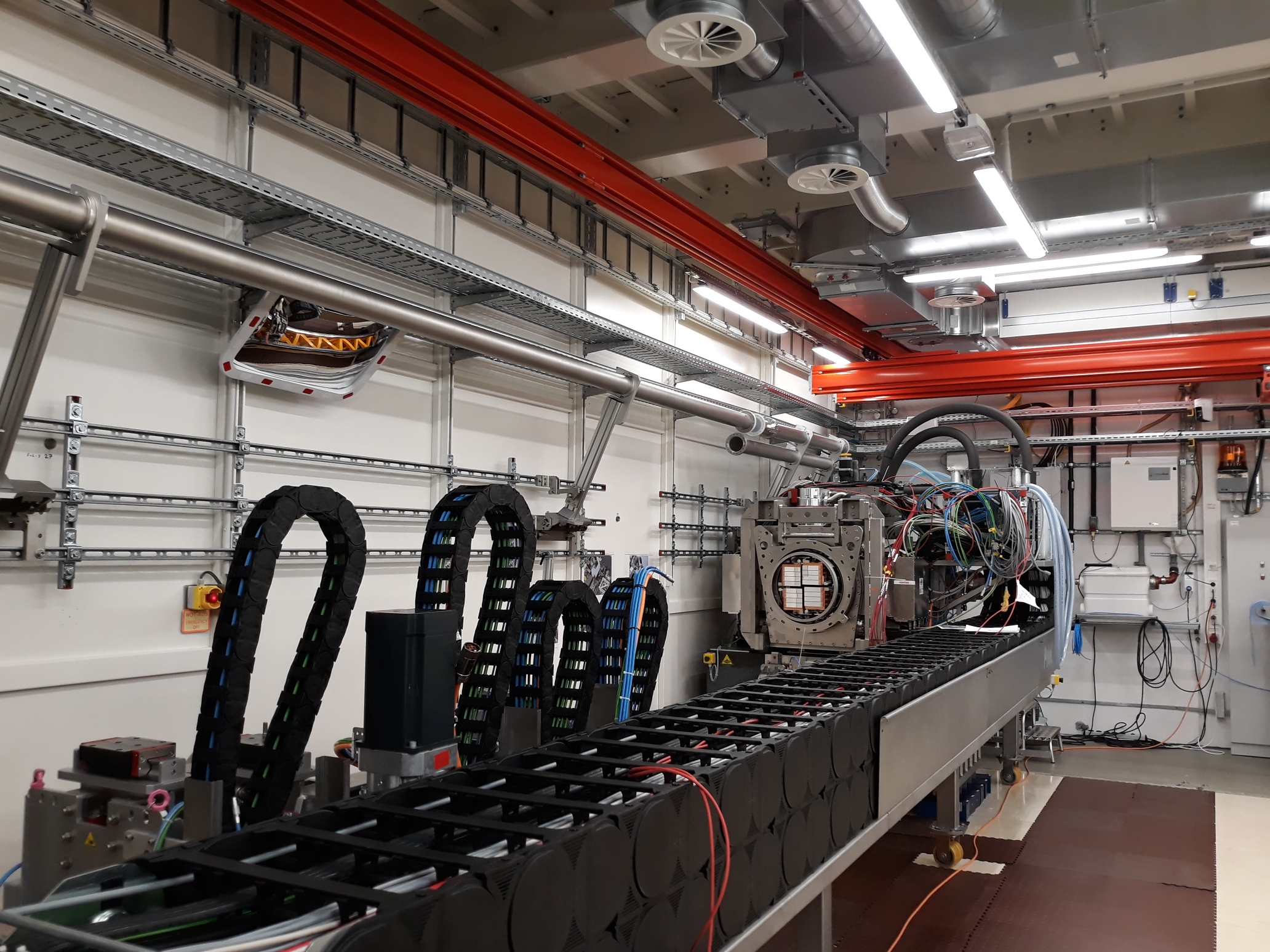} 
\caption{The AGIPD \SI{1}{\mega pixel} camera at the MID experimental station. The photograph shows the Detector at the end of the \SI{8}{\meter} long spectrometer arm, prior to the installation of the flight tube.}
\label{AGIPD_MID}
\end{figure}

The AGIPD \SI{1}{\mega pixel} camara at SPB has seen continuous user operation since the first user experiments in September 2017, while an identical camera has been delivered to the MID endstation (fig. \ref{AGIPD_MID}) in November 2018, which is currently being commissioned.

\section{First User Experiments with AGIPD}
\label{Sect:Firstexp}
\begin{figure*}[]
\centering
\includegraphics[width=\textwidth]{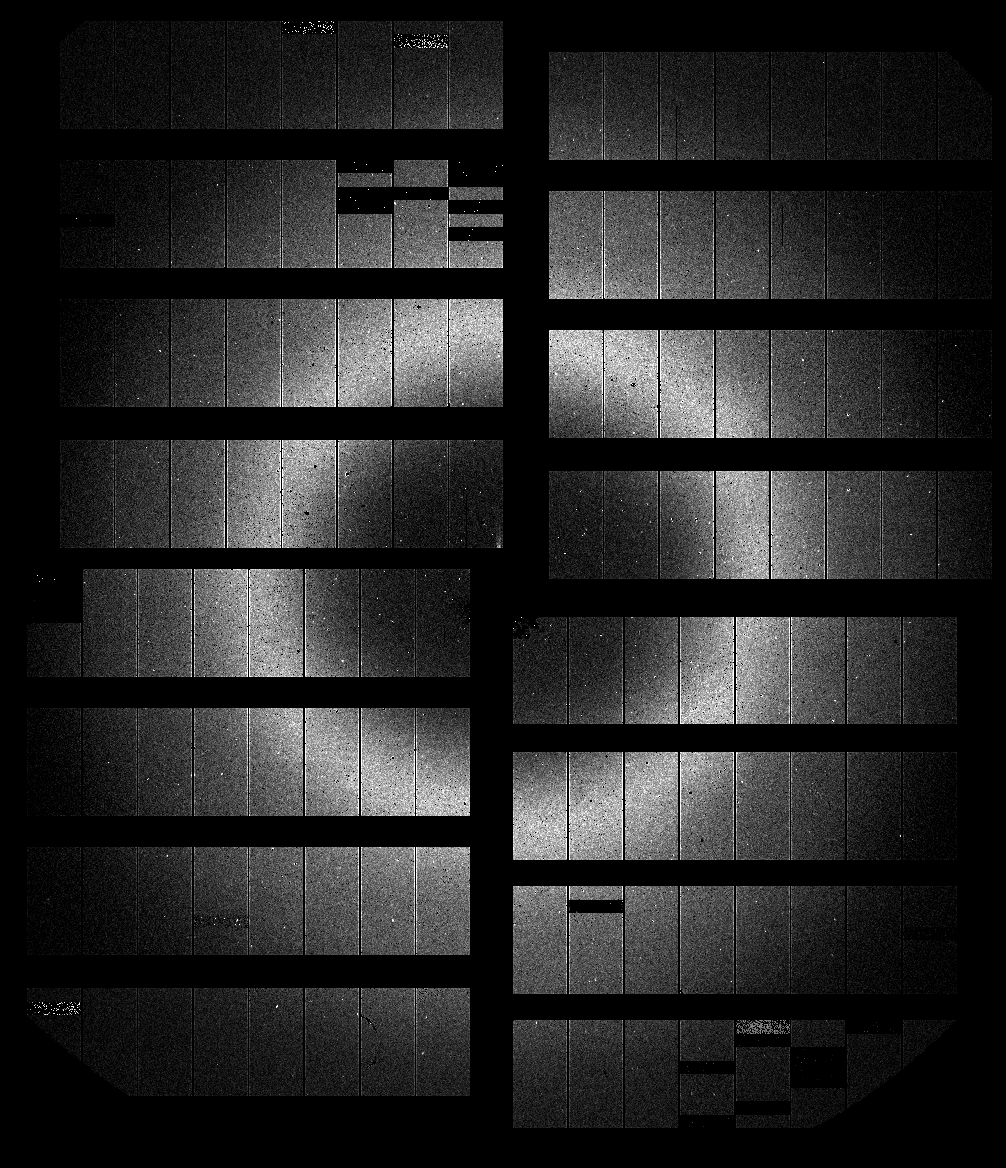} 
\caption{Diffraction pattern from one of the lysozyme micro crystals injected with a liquid jet. It is part of a \textsl{burst}, i.e. an image series recorded at \SI{1.1}{\mega\hertz} by the \SI{1}{\mega pixel} AGIPD at SPB. The diffraction spots and the \textsl{water ring} are nicely visible. The vertical stripes are masked double-sized pixels between readout ASICs.}
\label{Lyso}
\end{figure*}

\begin{figure*}[!tb]
\centering
\includegraphics[width=\textwidth]{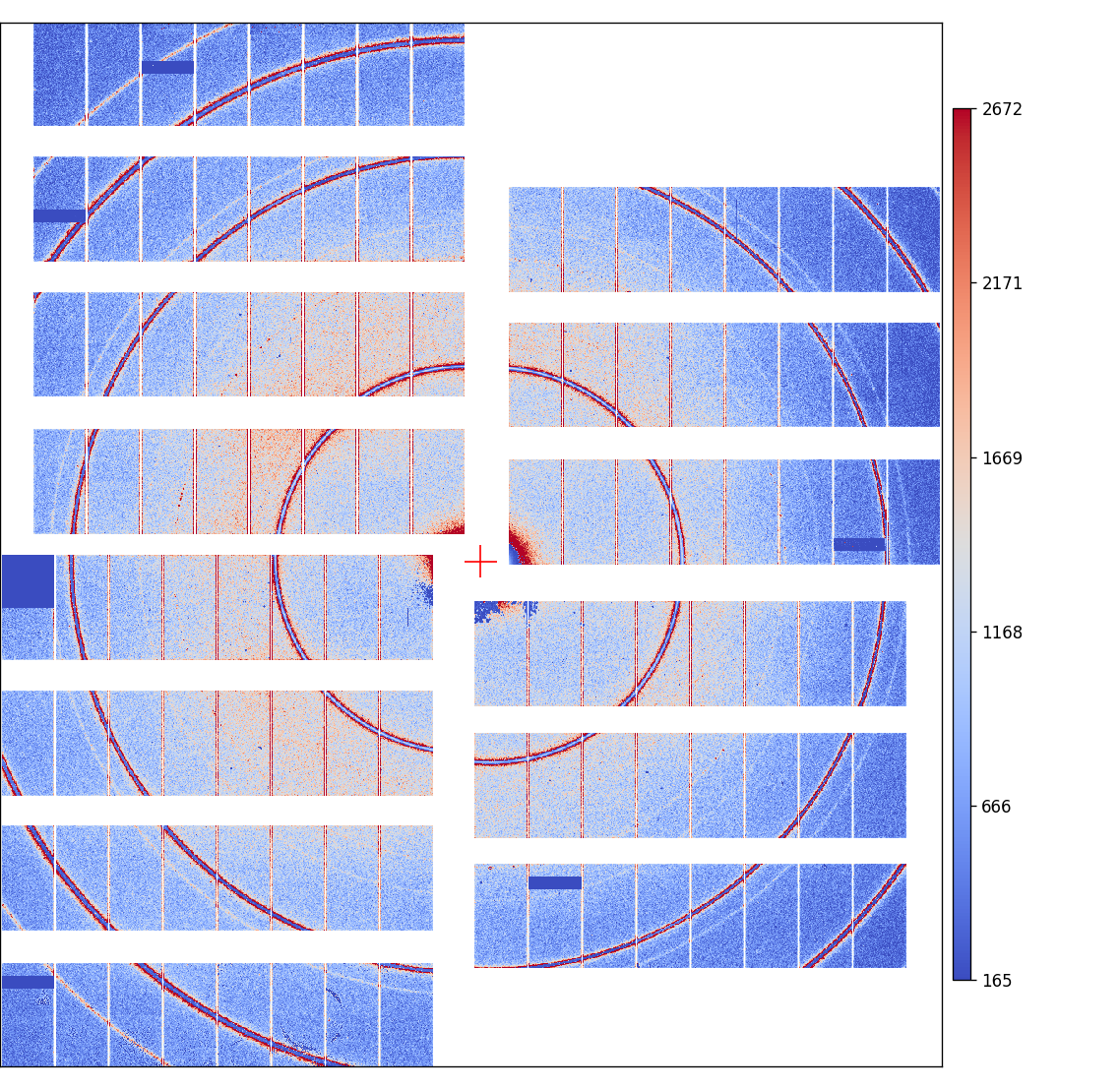} 
\caption{Powder diffraction pattern of Lithium Titanate ($\rm Li_2TiO_3$). It shows the uncorrected amplitude data recorded by the \SI{1}{\mega pixel} AGIPD at SPB. Gain switching is visible as lower amplitudes (blue and white colour in between red lines) in the most intense parts of the diffraction rings and some of the double-sized pixels (vertical stripes) between individual readout ASICs. }
\label{LiTi}
\end{figure*}

Due to its unique sampling rate, radiation hardness and dynamic range characteristics, AGIPD has been employed for experiments already in its prototyping stage and was e.g. used for an experiment to determine the coherence of the beam at PETRA III \citep{7}. 
The AGIPD \SI{1}{\mega pixel} system at the SPB station of the European XFEL was successfully demonstrated during the facility's inauguration and has seen continuous use since the very first user experiments in September 2017. In that context already the very first user experiment facilitated MHz serial femtosecond crystallography (at \SI{1.1}{\mega\hertz}, the rate of the accelerator during commissioning) for the first time and resolved the structure of CTX-M-14 \textbeta -lactamase for the first time \citep{XFEL-2012}. The figures \ref{Lyso} and \ref{LiTi} show a diffraction pattern from Lysozyme also recorded during that experiment and powder diffraction rings from Lithium-Titanate which are used for the spatial calibration, including the tilt of individual sensor tiles with respect to the detector plane.

\section{New AGIPD Detector Systems for SFX and HIBEF}
\label{Sect:AGIPDsfx}
For the SFX\footnote{\underline{S}erial \underline{F}emtosecond \underline{Crystal}lography} instrument at the European XFEL a \SI{4}{\mega pixel} AGIPD system is currently under construction. The image plane of this detector will consist of 56 sensors, arranged in a $4 \times 14$ pattern, formed by two halves with a vertical gap. These halves can be translated in the horizontal plane. The range lateral with respect to the beam is \SI{30}{\milli\meter}, while longitudinally the image plane can be moved by \SI{400}{\milli\meter} upstream -- through a giant gate valve of \SI{800}{\milli\meter} diameter into the experimental chamber. Unlike the \SI{1}{\mega pixel} systems, this system will use double FEMs, consisting of two LTCCs bolted to a cooled interposer. To cater the \textsl{side-by-side} arrangement of the sensors, which is incompatible with the arrangement of the readout electronics of the existing AGIPD \SI{1}{\mega pixel} systems at SPB and MID, and to implement lessons learned from these systems, a new readout board was designed.
\subsection{New readout boards}
\label{Subsect:ROB}
\begin{figure}[!tb]
\centering
\includegraphics[width=\columnwidth]{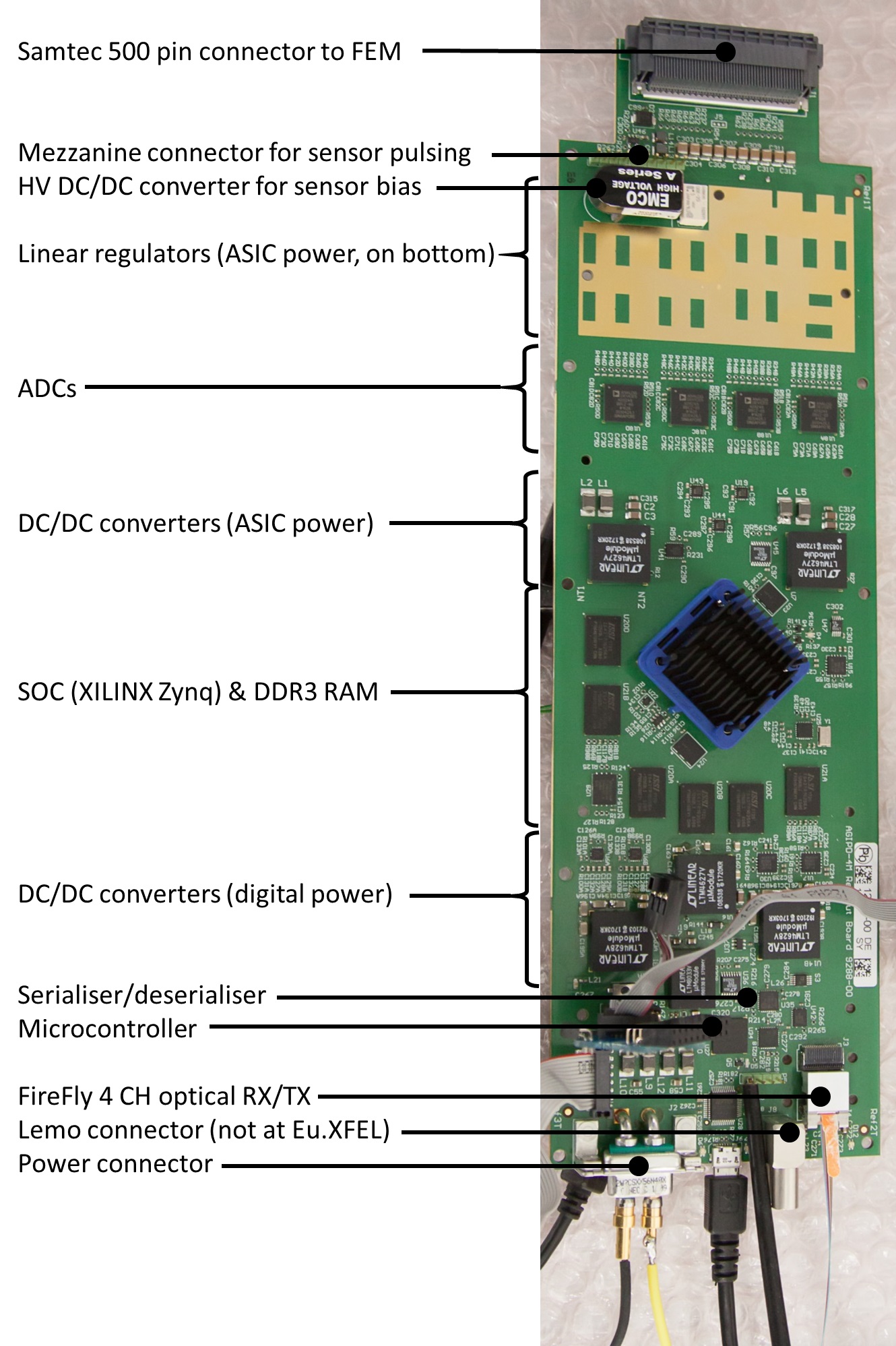} 
\caption{Photograph of a readout board used in the AGIPD systems for the SFX and HIBEF endstations.}
\label{newrob}
\end{figure}

During comissioning of the AGIPD \SI{1}{\mega pixel} system at SPB the vast number of connectors imposed a reliability problem, which required the external housings to be opened and the boards re-seated for proper contact. The same held, albeit to a lesser extent, true for the power supply cables, which also make moving the detector a major endeavour. A further shortcoming was the central generation of the ASIC commands by two \textsl{Master FPGA} boards, which led to signal integrity issues and a limited tunability of the ADC sampling phase. In turn the basic idea of the new readout board was the elimination of connectors and power supplies and a total modularity of the system. Thus the board houses all components to operate a  (single) FEM as a \textsl{stand-alone} detector (see fig. \ref{newrob}). Therefore, it implements the following components:
\begin{compactitem}
\item A HV DC/DC module to generate the sensor bias.
\item A connector for a mezzanine board to implement backside pulsing of the sensor as an additional means of calibration \citep{BSP}.
\item 14-bit ADCs directly driven by the AGIPD ASICs.
\item Cascaded switching and linear regulators to power the FEM's ASICs
\item A Xilinx\texttrademark\ Zynq\texttrademark\ SoC\footnote{System-on-chip} with DDR3 memory to control the ASICs, process the ADC data and send it on via a \SI{10}{GE} link.
\item A microcontroller to implement secondary tasks like power sequencing, configuration and monitoring.
\item A multimedia serialiser and deserialiser to interface with the European XFEL's clock and control (C\&C) and interlock systems.
\item A single power connector and DC/DC converters to power the board.
\item A Samtec\texttrademark\ FireFly\texttrademark\ \citep{FireFly} 4-channel optical transceiver for communications.
\item Two Lemo\texttrademark\ sockets for triggered stand-alone operation of an FEM without C\&C signals.
\end{compactitem}
The four channels of the FireFly transceiver are allocated to the \SI{10}{GE} data transmission, \SI{1}{GE} control interface, the serialised signals of the C\&C and interlock systems and an ethernet channel for debugging, which is not used during normal operation. 
Additional FireFly transceivers on the inside and outside of a vacuum flange re-group these signals such, that complete 4-channel ribbons with MTP connectors interface to the \SI{10}{GE} DAQ system, the \SI{1}{GE} control system and the \textsl{receiver board}, implementing the serialiser/deserialiser interface to the European XFEL's C\&C and interlock systems.

Since the readout boards are mounted within the vacuum vessel, a novell cooling concept is explored: Pairs of boards are attached to a liquid cooled heat exchanger by means of a \textsl{'Gap Filler'} thermal conductive plastic. \\

For the HIBEF\footnote{\underline{H}elmholtz \underline{I}nternational \underline{B}eamline for \underline{E}xtreme \underline{F}ields} endstation a \SI{1}{\mega pixel} system, based on the new readout board is under construction. The image plane of this detector will be a fixed \textsl{stack} of eight double modules, with a small gap between the top and bottom four, to allow the direct beam to pass. For this detector's  image plane only a translation along the beam axis, i.e. to adjust the distance of the detector from the sample, is foreseen. Translations in the other 2 dimensions are accomplished by the \textsl{detector bench}, i.e. by moving the detector's vacuum chamber. The biggest challenges for the HIBEF AGIPD detector system are the presence of pulsed high magnetic and electrostatic fields and photon energies $\ge \SI{25}{\kilo\electronvolt}$. At such energies the silicon sensor of AGIPD becomes \textsl{transparent}, i.e. inefficient and the usage of a high-Z sensor material like GaAs mandatory. 

\section{ecAGIPD - An electron collecting AGIPD ASIC for HIBEF}
\label{Sect:ecAGIPD}

Charge carrier lifetime -- especially of holes -- in high-Z semiconductor materials like GaAs and CdTe is short compared to elementary semiconductors like Si or Ge \cite{Owens}. Recent advancements in the production of these sensor materials \cite{Veale} mitigate effects like '\textsl{afterglow}' and '\textsl{polarisation}' (decribed in \cite{Siemens}) and made compound semiconductor sensors the subject of investigations for an alternative to Germanium sensors in high-flux -- high-energy imaging detectors at FELs \cite{Becker_CdTe, Becker_GaAs, Veale_GaAs}. In addition sensors made from high-Z materials (including Germanium) do not show the defect mechanisms of silicon decribed in \cite{Jiaguo_UHH}, while the high absorption of radiation in the sensor and the low cross section of silicon at high photon energies drastically reduces the dose deposited in the readout ASICs. 
However the current AGIPD ASICs (and sensors) are \textsl{hole collecting} devices, and thus not suitable for high-Z materials. For this reason an electron collecting version of the AGIPD ASIC, ecAGIPD, to equip the AGIPD camera at the HIBEF endstation with high-Z sensors is being developed. \\

\begin{figure*}[!htb]
\centering
\includegraphics[width=\textwidth]{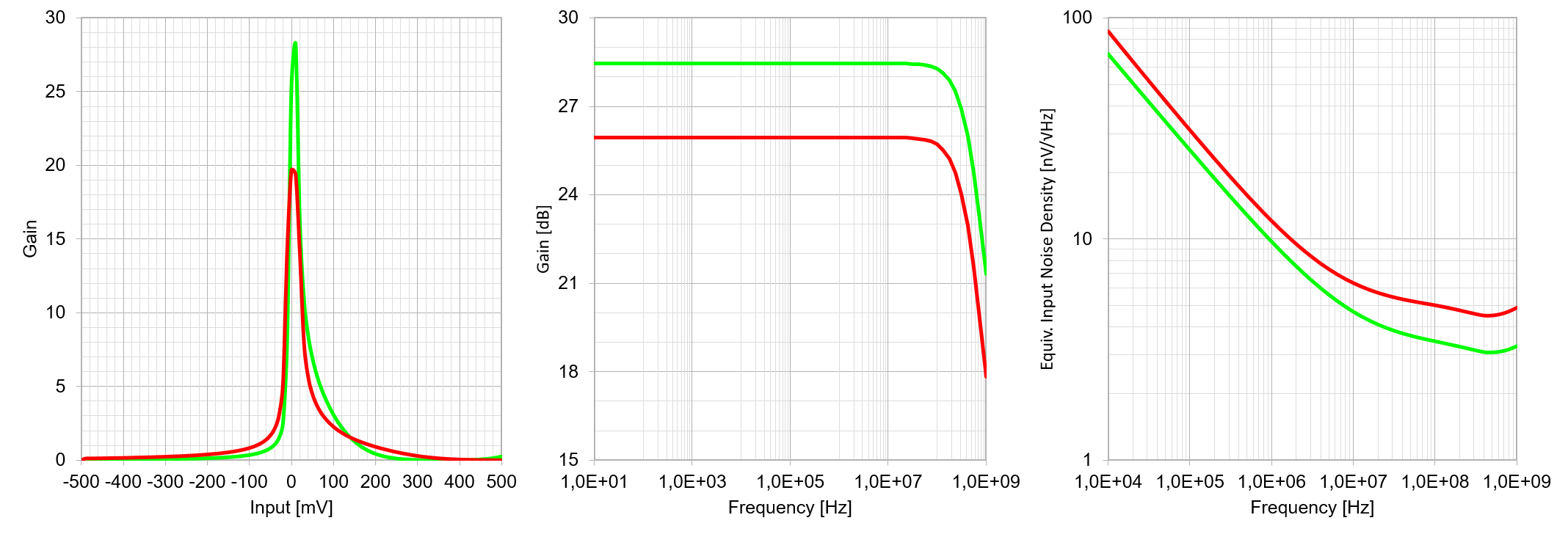} 
\caption{\textsl{Spectre} Simulation results showing (from left to right) gain ($\times 29$ compared to $\times 19$), frequency response (\SI{-3}{\decibel} at $\approx \SI{500}{\mega\hertz}$) and noise density of the preamplifier core of the electron collecting ecAGIPD ASIC (green) in comparison to the hole collecting AGIPD (red).}
\label{ecSIM}
\end{figure*}

\begin{figure*}[!htp]
\includegraphics[width=\textwidth]{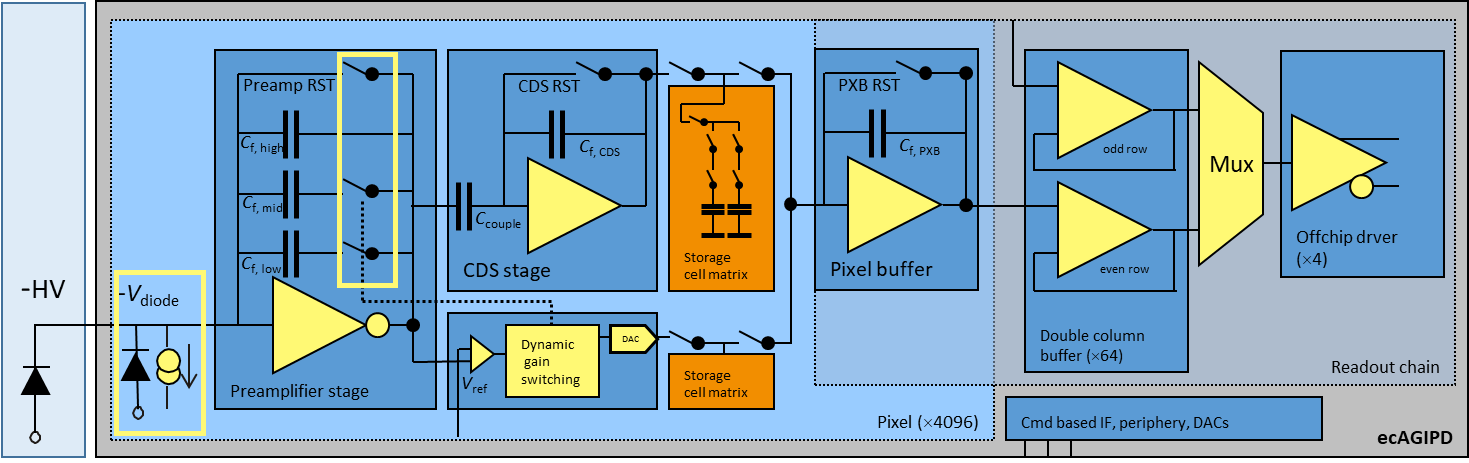}
\caption{Circuit schematic of the ecAGIPD, the electron collecting version of the AGIPD ASIC. The yellow frames contain the components located in isolated p-wells.}
\label{ecAGIPD_schema}
\end{figure*}

The primary difference is the lowered baseline (operating point at $\approx \SI{400}{\milli\volt}$) of the charge sensitive preamplifier. This is necessary to maintain a dynamic range of $\approx \num{e4}$ \SI{12.4}{\kilo\electronvolt} photons towards the positive supply rail. Since the baseline is only minimally higher than an n-MOS' threshold voltage, it requires the protection of the amplifier inputs and calibration stimuli sources to work from a negative potential, which was implemented using the \textsl{triple well} option of the GF \SI{130}{\nano\meter} process. These p-wells are biased below substrate potential to allow bias sources, input protection diodes, and n-MOS feedback switches to operate properly \- even in presence of large input charges. By these changes the performance of the preamplifier core actually improves with respect to the hole collecting version, as the simulation results in fig. \ref{ecSIM} predict.

Further changes to the circuit are minor: Besides the obvious reversal of the discriminator's polarity required for gain switching, also the levels encoding high and low gain and the pads of the differential analogue outputs were swapped. This way ecAGIPD will deliver the same signal polarities to the subsequent electronics \- facilitating the reuse of the existing firmware and calibration algorithms. The block schematic is shown in fig. \ref{ecAGIPD_schema}.
To evaluate the design AGIPD 0.6, a $16 \times 16$ pixels prototype of ecAGIPD, has been manufactured and is currently characterised.

\section{Going faster}
\label{Sect:GoingFaster}
European XFEL plans to introduce two additional bunch patterns in the $\rm 2^{nd}$ half of the 2020ies \cite{CW_LP}. These are foreseen to be 
\begin{compactitem}
\item CW operation at \SI{100}{\kilo\hertz} rate
\item \textsl{Long pulse mode} with \SI{500}{\milli\second} long bursts of $\le \SI{200}{\kilo\hertz}$ pulses
\end{compactitem}
and would result in an $\times 3.7$ increase in brilliance, if the pulse parameters are kept\footnote{Other limits, like e.g. heat load of superconducting accelerator parts might prevent an increase in brilliance.}. The largest benefit of the CW mode is the ability exchange (or at least move) samples in-between pulses and this way reduce radiation damage to the sample and enable pulse-by-pulse experiments beyond jet based sample delivery.

The burst frame rate of an integrating detector like AGIPD is in principle only limited by the properties of sensor and preamplifier. For the future CW operation of the European XFEL, readout bandwidth will become a bottleneck. The AGIPD ASIC can cope with CW operation up to a  frame rate of $\rm \approx \SI{16}{\kilo\hertz}$ with original performance \footnote{This mode is used during the characterisation and wafer-level testing of single ASICs. The readout systems of the SPB and MID AGIPD cameras require a network and firmware upgrade to implement it.}. At even higher rates (theoretically the ASIC can work at up to \SI{96}{\kilo\hertz} when omitting gain readout), the deterioration of the analogue readout signals due to e.g. skin effect\footnote{At frequencies of several \SI{10}{\mega\hertz}, signal currents will gradually start to flow only on the surface of a conductor, leading to a rise of impedance and a non-flat frequency response.}, and reflections\footnote{Reflections, i.e. parts of a signal traveling an electrical transmission line in opposite direction, are generated at any discontinuity of impedance. This is predominantly visible for higher frequencies at e.g. connectors and non-matching termination resistors.} will render operation impractical.

\begin{figure}[!tb]
\centering
\includegraphics[width=\columnwidth]{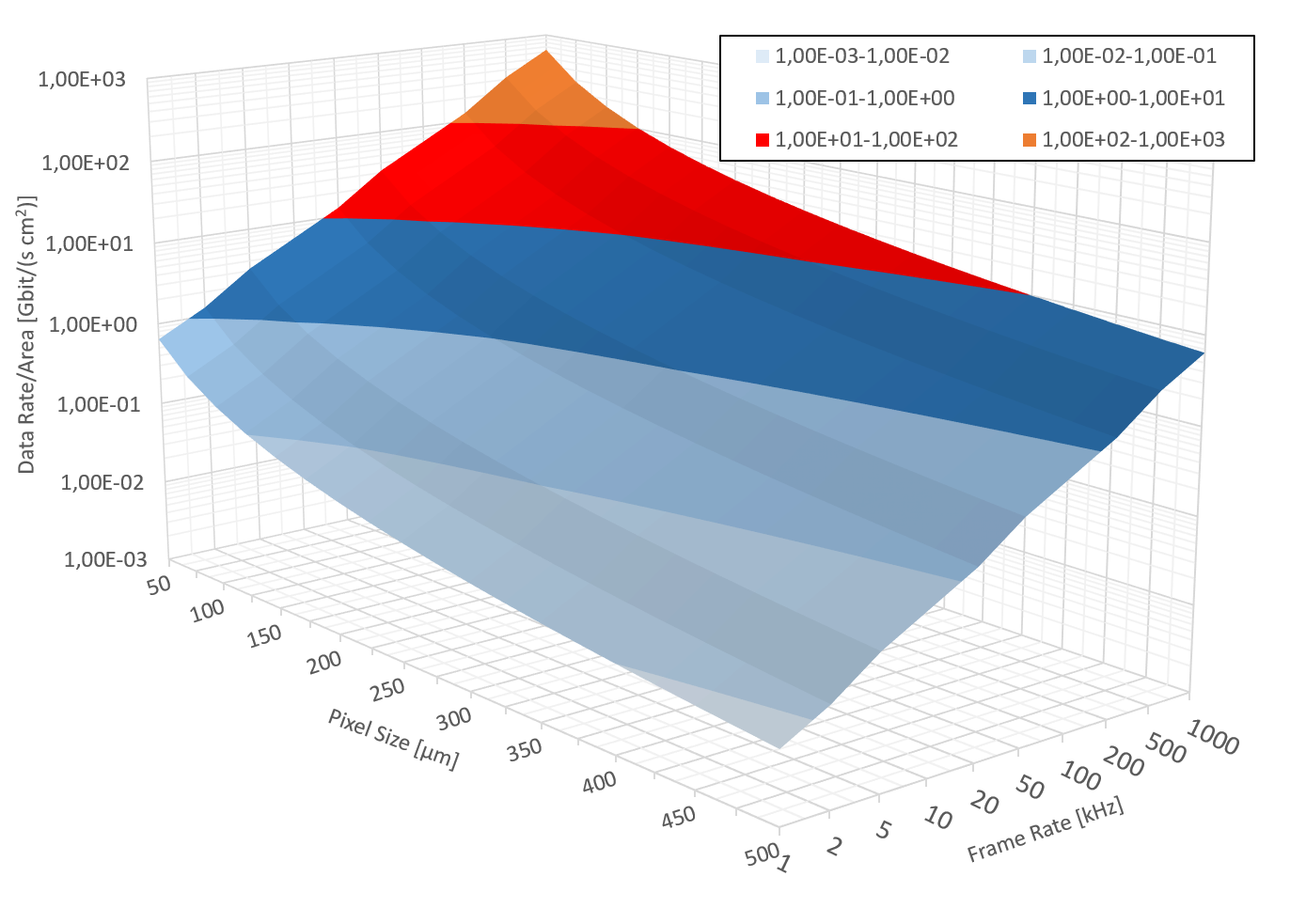} 
\caption{Data rate per area as a function of pixel size and frame speed, assuming a \SI{16}{bit} quantisation.}
\label{pix_rate}
\end{figure}

As a consequence, in-pixel digitisation becomes a must, while readout bandwidth remains one of the limiting factors. Assuming current technology (FireFly with \SI{28}{\giga bit \per\second \per{link}}), data rates of about 
\SI{48}{\giga bit \per\second\per\square{\centi\meter}} or 
\SI{3}{\giga s \per\second\per\square{\centi\meter}} at 
\SI{16}{bit\per{sample}} can be reached. Fig. \ref{pix_rate} shows data rate as a function of pixel size and frame speed under these conditions.  

An even more challenging problem for such detectors will be power dissipation, since dynamic gain switching requires the the preamplifier to be faster than the sensor's charge collection time. Hence power consumption is inversely proportional to the pixel area and will not relax with the lower frame rate compared to the current \SI{4.5}{\mega\hertz} detectors at the European XFEL. In turn pixels smaller than $\approx \SI{100}{\micro\meter} \times\SI{100}{\micro\meter}$ will result in a power dissipation of $\gtrapprox \SI{4}{\watt\per\centi\meter\square}$ and will be difficult to implement.

\section{Summary}
A \SI{1}{megapixel} AGIPD detector system has been installed at the SPB instrument of the European XFEL in August 2017.
System fulfils all requirements, esp. in terms of noise, which is below \SI{310}{\electron} (i.e. $\rm \le \SI{1.2}{\kilo\electronvolt}$), single photon sensitivity, dynamic range ($\rm \ge \SI{e4}{photons}$ at \SI{12}{\kilo\electronvolt}) and speed(\SI{4.5}{\mega\hertz} frame rate). This system has been extensively used in user operation since day one.
Remaining shortcomings, like an increasing uncertainty of the gain determination during readout or a limited number of frames per burst are tackled by a new version of the readout ASIC (AGIPD 1.2, taped out in Aug. 2018) and ongoing firmware development.\\
An identical \SI{1}{megapixel} system has been delivered to the MID endstation in November 2018 and is currently being commissioned.

AGIPD systems of \SI{4}{megapixel} for the SFX endstation and of \SI{1}{megapixel} for HIBEF are currently under construction. These systems are based on a new readout board, which implements autonomous operation of each detector module from a single power supply and data transmission and communications purely via optical fibres. Since prototypes showed no issues, production of these boards has started.
For these boards and for the sensor modules novel in-vacuum cooling concepts are evaluated.
The AGIPD \SI{1}{megapixel} system for HIBEF will be equipped with sensors made from high-Z semiconductor material, for which an electron collecting version of the AGIPD ASIC (ecAGIPD) is being developed. A $\rm 16 \times \SI{16}{pixel}$ demonstrator has been manufactured and is awaiting evaluation.
Since European XFEL will provide different pulse patterns in the $\rm 2^{nd}$ half of the 2020ies, we are studying concepts for an ultra high framerate ($\rm \ge \SI{100}{\kilo\hertz}$) imager with in-pixel digitisation, where power consumption and readout bandwidth become limiting factors.






\end{document}